%% file: main.tex
\documentclass[sigplan,nonacm]{acmart}
\usepackage[linesnumbered,ruled,vlined,noend]{algorithm2e}
\DontPrintSemicolon

\SetKwComment{Comment}{\color{green!50!black}// }{}

\newcommand{\assign}{\leftarrow}
\newcommand{\var}{\texttt}
\newcommand{\typehint}{\textbf}
\newcommand{\FuncCall}[2]{\texttt{\bfseries #1(#2)}}
\SetKwProg{Function}{function}{}{}
\SetKw{kwMax}{max}
\SetKw{KwBreak}{break}

\graphicspath{ {./src/} }

\definecolor{Gray}{gray}{0.9}

\definecolor{theme-green}{HTML}{89b374}
\definecolor{theme-darkblue}{HTML}{384890}
\definecolor{theme-grey}{HTML}{e4e4e4}
\definecolor{theme-darkgrey}{HTML}{808080}
\definecolor{theme-blue}{HTML}{387490}
\definecolor{theme-purple}{HTML}{903874}

\usepackage{accsupp}
\newcommand{\noncopynumber}[1]{%
	\BeginAccSupp{method=escape,ActualText={}}%
	\tiny\color{theme-darkgrey}
    #1%
	\EndAccSupp{}%
}

\usepackage[]{hyperref}

\usepackage{colortbl}
\usepackage{titlesec}
\usepackage{booktabs}
\usepackage{graphics}
\usepackage{tabularx}
\usepackage{arydshln}
\PassOptionsToPackage{table}{xcolor}
\usepackage{graphicx}
\usepackage{makecell}
\usepackage{enumitem}
\setlist[enumerate,1]{nosep, leftmargin=*}
\setlist[itemize,1]{nosep, leftmargin=*}

\usepackage{afterpage}

\usepackage{subcaption}
\usepackage{caption}

\usepackage{stfloats}

\usepackage{listings}

\DeclareCaptionFormat{listing}{}

\author{Alban Dutilleul}
\affiliation{\institution{ENS Rennes} \country{France}}
\email{alban.dutilleul@ens-rennes.fr}

\author{Hugo Pompougnac}
\affiliation{\institution{Univ. Grenoble Alpes, Inria, CNRS, Grenoble INP, LIG, 38000 Grenoble, France} \country{}}
\email{hugo.pompougnac@inria.fr}

\author{Nicolas Derumigny}
\affiliation{Télécom SudParis \country{France}}
\email{derumigny.nicolas@telecom-sudparis.eu}

\author{Gabriel Rodríguez}
\affiliation{\institution{Universidade da Coruña} \country{Spain}}
\email{gabriel.rodriguez@udc.es}

\author{Valentin Trophime}
\affiliation{\institution{Univ. Grenoble Alpes, Inria, CNRS, Grenoble INP, LIG, 38000 Grenoble, France} \country{}}
\email{valentin.trophime@inria.fr}

\author{Christophe Guillon}
\affiliation{\institution{Univ. Grenoble Alpes, Inria, CNRS, Grenoble INP, LIG, 38000 Grenoble, France} \country{}}
\email{christophe.guillon@inria.fr}

\author{Fabrice Rastello}
\affiliation{\institution{Univ. Grenoble Alpes, Inria, CNRS, Grenoble INP, LIG, 38000 Grenoble, France} \country{}}
\email{fabrice.rastello@inria.fr}

\input{defs}

\title{Performance Debugging through Microarchitectural Sensitivity and Causality Analysis}

\begin{document}
%-------------------------------------------------------------------------------

\date{}

\input{src/00-abstract.tex}

\maketitle

\input{src/01-introduction.tex}
\input{src/02-bottleneckness.tex}
\input{src/03-gus.tex}
\input{src/04-evaluation.tex}

\input{src/05-related-works.tex}
\input{src/06-conclusion.tex}

%-------------------------------------------------------------------------------
\bibliographystyle{plain}
\bibliography{main}

%%%%%%%%%%%%%%%%%%%%%%%%%%%%%%%%%%%%%%%%%%%%%%%%%%%%%%%%%%%%%%%%%%%%%%%%%%%%%%%%
\end{document}

%% file: defs.tex
\def\taint{\var{taint}\xspace}

\def\dispatchqueue{\var{dispatchqueue}\xspace}
\def\taintqueue{\var{taintqueue}\xspace}
\def\dispatch{\var{dispatch}\xspace}
\def\frontend{\var{frontend}\xspace}
\def\ithr{\var{inversethroughput}\xspace}
\def\i{\var{i}\xspace}
\def\T{\var{T}\xspace}
\def\loc{\var{loc}\xspace}
\def\res{\var{res}\xspace}
\def\Ll{\var{L}_\var{i}\xspace}
\def\this{\var{this}\xspace}
\def\c{\var{c}\xspace}
\def\tdispatch{t_{\dispatch}}

\newcommand{\coz}{\texttt{Coz}}
\def\gus{\textsc{Gus}\xspace}
\newcommand{\toolsensitivity}{\gus}
\newcommand{\qemu}{\texttt{QEMU}}

\newcommand{\uopsinfo}{\texttt{UOPS.INFO}}

\newcommand{\tstart}{\texttt{t\textsubscript{start}}}
\newcommand{\tend}{\texttt{t\textsubscript{end}}}
\def\tmin{\texttt{t\textsubscript{min}}\xspace}

\def\tavail{\texttt{t\textsubscript{avail}}\xspace}

\def\llvmmca{\texttt{llvm-mca}\xspace}

\def\iaca{\texttt{iaca}\xspace}
\newcommand{\vtune}{\texttt{VTune}}
\newcommand{\perf}{\texttt{perf}}

\def\facile{\texttt{facile}\xspace}
\def\uica{\texttt{uica}\xspace}
\newcommand{\osaca}{\texttt{osaca}} 
\newcommand{\granite}{\texttt{granite}}
\newcommand{\ithemal}{\texttt{ithemal}}

\newcommand{\ie}{\textit{i.e.}}
\newcommand{\etc}{\textit{etc.}}

\newcommand{\muop}[1]{$\mu$op#1}

\newcommand{\figref}[1]{Fig.~\ref{fig:#1}}

\newcommand{\decan}{\texttt{DECAN}}
\newcommand{\maqao}{\texttt{MAQAO}}
\newcommand{\saake}{\texttt{SAAKE}}

%% file: src/00-abstract.tex
%-------------------------------------------------------------------------------
\begin{abstract}
%-------------------------------------------------------------------------------

Modern Out-of-Order (\textit{OoO}) CPUs are complex systems with many components interleaved in non-trivial ways. 
Pinpointing performance bottlenecks and understanding the underlying causes of program performance issues are critical tasks to fully exploit the performance offered by hardware resources.

Current performance debugging approaches rely either on measuring resource utilization, in order to estimate which parts of a CPU induce performance limitations, or on code-based analysis deriving bottleneck information from capacity/throughput models. 
These approaches are limited by instrumental and methodological precision, present portability constraints across different microarchitectures, and often offer factual information about resource constraints, but not causal hints about how to solve them.

This paper presents a novel performance debugging and analysis tool that implements a resource-centric CPU model driven by dynamic binary instrumentation that is capable of detecting complex bottlenecks caused by an interplay of hardware and software factors. 
Bottlenecks are detected through \textbf{sensitivity-based analysis}, a sort of model parameterization that uses differential analysis to reveal constrained resources.
It also implements a new technique we developed that we call \textbf{causality analysis}, that propagates constraints to pinpoint how each instruction contribute to the overall execution time.

To evaluate our analysis tool, we considered the set of high-performance computing kernels obtained by applying a wide range of transformations from the Polybench benchmark suite~\cite{polybench} and measured the precision on a few Intel CPU and Arm micro-architectures.
We also took one of the benchmarks (\texttt{correlation}) as an illustrative example to illustrate how our tool's bottleneck analysis can be used to optimize a code.

\end{abstract}

%% file: src/01-introduction.tex
%-------------------------------------------------------------------------------
\section{Introduction}\label{sec:intro}
%-------------------------------------------------------------------------------

In the last decades, high-performance computing has become one of the fundamental pillars of scientific innovation. 
The end of Dennard scaling in the early 21st century fundamentally shifted the architectural approach to performance improvements, which had been based on increasing clock frequencies with constant MOSFET power densities, and into a ``dark silicon''~\cite{darksillicon2011} era in which systems are composed of a heterogeneous collection of computational devices with very different performance-power operating points.

%FIXME: hook between the previous paragraph and the focus on CPUs on the next? We are waiving away GPUs, etc.

While the basics of modern processor designs closely follow the ideas by Robert Tomasulo, debuted in the IBM 360/91 in 1966, CPU cores are nowadays more complex machines. 
In 2006, Intel launched its "Core" architecture, featuring a 65nm process node, a 4-uops issue width, a 96-entry ROB, and a 32-entry unified schedulers that feeds 6 superscalar ports. 
In 2023, the Meteor Lake architecture features both performance- and efficient-oriented processors, as well as an integrated GPU. 
The process node varies for different parts of the chip. 
The 7nm Redwood Cove P-processor has a 6-uops issue width, with three different instruction-decoding paths, a 512-entry ROB and 3 different scheduling units with a combined 215 entries that feed a 12-port superscalar architecture. 
Add to the mix Simultaneous Multi-Threading (SMT) support, and the result is an
extraordinarily complex architecture, on which \emph{performance debugging} becomes a
daunting task.

A popular way to analyze the internal behavior of a CPU for a specific computational kernel is to rely on the per\-for\-mance-monitoring counters (PMCs) made available by the performance-monitoring unit (PMU) of modern CPUs. 
These model-specific registers provide a native way to measure the frequency of different low-level events in the CPU, offering a glimpse at the internal workings of this otherwise black box. 
The state-of-the-art \emph{Top-down Microarchitecture Analysis} (TMA) method~\cite{topdown} defines a methodology in which low-level events are measured and used to compute aggregation formulas that pinpoint the areas of the processor that become performance bottlenecks during the execution of a given computational kernel. 
While TMA provides invaluable leads to guide optimization, it has limitations stemming both from precision, methodological, and portability concerns. 
\emph{Binary code analysis}~\cite{iaca,llvmmca,uica} is a complementary approach based on a performance model at the assembly level. 
Although it produces less precise estimations than PMC-based measurements, assembly models a wider range of metrics and greater flexibility in analysis, while not requiring any hardware support.

This work proposes two novel performance-debugging methods respectivelly called \emph{causality} and \emph{sensitivity analysis}. 
It is based on a coarse-level model-based simulation, implemented through instrumentation of a binary. 
This simulation is used to power:
(1) a constraints propagation engine, that allows to pinpoint how each individual static instruction contributes to the overall execution time;
(2) a differential analysis engine, that studies the performance effects of varying the capacities of different CPU resources, identifying both the CPU functional blocks that act as bottlenecks during the execution of a particular kernel, as well as the bottlenecked sections of code on which to focus the optimization efforts. 
The contributions of this work are as follows:

\begin{itemize}
\item A novel causality-based approach to performance debugging which provides information about which specific instruction contribute to the overall execution time.
	\item A novel sensitivity-based approach to performance debugging which provides information about the specific parts of the CPU which limit the performance of a given computational kernel.
	\item An end-to-end implementation of \gus, a sensitivity-based profiler for performance debugging.
	\item An extensive experimental validation over 1624 kernels on 6 different
          microarchitectures, demonstrating the accuracy
          of \gus compared to measurements.
        \item A case study illustrating the benefit of our bottleneck analyzer for performance debugging.
\end{itemize}

The paper is organized as follows. Section~\ref{sec:bottleneckness} gives a comprehensive description of existing performance analysis techniques and motivates our approach. Section~\ref{sec:gus} describes \gus. Section~\ref{sec:experiments} reports our experimental results. Section~\ref{sec:discussion} discussed our implementation choices. Section~\ref{sec:conclusion} concludes.

%% file: src/02-bottleneckness.tex
%-------------------------------------------------------------------------------
\section{Performance debugging: state of the art and motivation}\label{sec:bottleneckness}
%-------------------------------------------------------------------------------

Consider the assembly kernel shown in Fig.~\ref{lst:Motivation:Running-example}(a), that yields a raw performance of 12.8~GFLOPS when executed on an Intel i9-12900k core with a theoretical peak performance of 153.6~GFLOPS\footnotemark. 
Figuring out where the \emph{bottleneck} of this computation lies for the given microarchitecture is a remarkably complex task. 
Here, a bottleneck is defined as a microarchitectural resource or characteristic that becomes a fundamental cause of performance degradation due to the limitations it imposes to the flow of the computation. 
On the one hand, one could hypothesize that this kernel is memory-bound due to the \texttt{vmovaps} in line 3. 
However, depending on the access pattern and the prefetcher this load might be served from L1. 
If so, and depending on the arithmetic intensity of the kernel, the FP FMA units may become saturated and act as the computational bottleneck. 
This is the fundamental trade-off studied by the Roofline model~\cite{williams2009roofline}, which predicts the expected performance of a kernel from its arithmetic intensity and the architectural parameters of the computer.

\footnotetext{This processor features 3xFP-FMA AVX-2 units running at a base frequency of 3.2~GHz. These measurements refer to single-precision FLOPS.}

\input{src/running-example}

Roofline analysis models a simplified, ideal version of the processor, useful to assess whether a computation achieves peak performance. 
However, when that does not happen, it does not provide any information as to \emph{why}. 
In a real processor, different parts can become a performance bottleneck, depending on the computation itself, its implementation, the microarchitectural details, and even the execution context (e.g., load by other tasks, operational temperature, etc.). 
For example, if the calculation of the addresses accessed by the \texttt{vmovaps} in line 3 is complex enough, it could saturate the integer ALUs and become a bottleneck itself.

\begin{figure}
	\begin{center}
		\includegraphics[width=\linewidth]{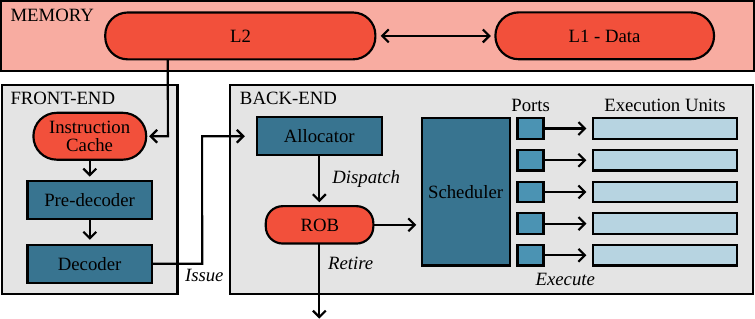}
	\end{center}
	\vspace{-1em}
	\caption{Simplified view of a pipelined OoO CPU core.}
	\label{fig:cpu}
	\vspace{-1em}
\end{figure}

\figref{cpu} shows a simplified, high-level representation of a pipelined out-of-order CPU core. 
During its lifecycle, each \muop{} passes through four distinct states: 
it is \emph{issued} by the front-end, \emph{dispatched} by the allocator, \emph{executed} by the functional units during its stay in the Re-Order Buffer (ROB), and eventually \emph{retired} once its effects have been committed. 
During this process, any of the following parts of the CPU may incur performance penalties, and ultimately become a performance bottleneck:

\paragraph{Front-end (F-E)}
This is the part of the processor in charge of fetching and decoding instructions from memory. 
It includes the Branch Prediction Unit (BPU), which guides the control flow by setting the value of the Program Counter (PC), potentially speculating on branch outcomes. 
Instructions are either read from the Instruction Cache (IC) and decoded into micro-operations (\muop{s}), or fetched from a \muop{}-cache. 
If the F-E is not capable of decoding \muop{s} at the rate that the processor executes them, it can become a bottleneck.

%\paragraph{Allocator/Renamer (AR)} The F-E issues \muop{s} for execution to the Back-end (B-E) through a set of \emph{pipeline slots}\footnotemark. Upon arrival, the AR assigns \muop{s} to an execution port~\cite{intel}. At the same time, register renaming takes place, translating each architectural register used by the \muop{} into a physical register. This operation, controlled by the Register Alias Table (RAT), removes write-after-read and write-after-write hazards, and allows to use the pool of physical registers in the core, which is typically much larger than the set of architectural registers defined by the ISA.
%{\color{red}[can we remove the AR / fuse the interesting parts with the ROB paragraph? I don't think that the AR can become a bottleneck, and while this paragraph is informative it is not like the others in the sense that I think it will not be referenced again]}

\paragraph{Re-Order Buffer (ROB)}\label{par:rob}
The F-E issues \muop{s} for execution to the Back-end (B-E) through a set of \emph{pipeline slots}\footnotemark. 
Upon arrival, the Allocator assigns \muop{s} to an execution port~\cite{intel}, and places them in the ROB, a circular buffer which allows for in-order termination with out-of-order execution, enabling simple recovery from branch mispredictions. 
When a \muop{} is \emph{retired}, its architectural effects are commited and it is removed from the ROB. 
When all ROB slots become full the processor is stalled, i.e., the issue of \muop{s} from the F-E to the B-E is stopped, until a slot becomes free.

\footnotetext{In this context, this term refers to each of the multiple paths between the F-E and the B-E. 
  In each cycle, the front-end sends either zero or one \muop{s} to the back-end for execution through each of these paths. 
  The number of pipeline slots is a microarchitectural parameter.}

\paragraph{Scheduler/Reservation Station (RS)}
Upon execution, \muop{s} are held in a RS, which includes a copy of their available input data, until its dependencies are satisfied and its assigned execution port becomes free. 
When these conditions are met, the \muop{} is effectively executed. 
When all the RS become full the processor must be stalled.

\paragraph{Execution ports and execution units}
Each port is linked to execution units responsible for a class of operations: 
integer ALUs, divider, memory accesses, vector computations, \etc{}. 
A port can process a maximum of one \muop{} per cycle. 
A lack of execution ports of a given type may cause a backlog, possibly leading to the ROB and/or RS becoming saturated and processor stalls.

\paragraph{Memory instructions}
Memory loads and stores are placed into Load and Store Buffers, respectively. 
The latency of Load-Store units is determined by the level of the memory hierarchy into which the retrieved data is located. 
A high memory latency may cause \muop{s} depending on loaded data to backlog, possibly inducing saturation of the ROB and/or RS and processor stalls.

\subsection{Performance debugging and optimization}
A state-of-the-art technique for characterizing and debugging the performance of a computational kernel is the Top-down Microarchitecture Analysis (TMA) method~\cite{topdown}, developed by Intel engineers and actively supported as part of Perfmon Metrics~\cite{Intel:PerfMon}. Conceptually, TMA studies what happens to each pipeline slot in each cycle of the execution of an application. If $I_w$ is the architectural issue width, $T_c$ is the execution time of the application, and $F$ is the average frequency, TMA categorizes each of the $I_w \cdot T_c \cdot F$ pipeline slots into one of four categories:

\begin{itemize}
	\item \emph{Retired}: this slot issued a \muop{} that was eventually retired. This indicates useful computation\footnotemark.
	\item \emph{Bad speculation}: this slot issued a \muop{} that was eventually \emph{not} retired, i.e., it was purged from the ROB without being committed. This is indicative of misspeculation.
	\item \emph{F-E bound}: this slot remained unused due to no decoded \muop{s} being available. This is indicative of F-E boundedness.
	\item \emph{B-E bound}: this slot remained unused due to the ROB/RS being full. This is indicative of B-E boundedness.
\end{itemize}

\footnotetext{But not necessarily \emph{efficient} computation. For instance, a non-vectorized computation might have a large fraction of its pipeline slots in the \emph{Retired} category, yet be far from peak performance.}

\noindent{}In an ideal situation, all the pipeline slots belong to the Retired category. 
In reality, the distribution of the slots among the four categories indicates how efficiently the CPU is being used, and which of its parts become bottlenecks. 
TMA defines a hierarchy of subcategories, further refining this top-level classification to give more insight about specific microarchitectural bottlenecks. 
For instance, a slot into the back-end bound category can be further classified into \emph{core bound} and \emph{memory bound} categories, and in turn a memory bound slot can be attributed to the latency of any of the levels of the memory hierarchy.

%{\color{red}
%[inserted from sec1, this will be useful]
%In a modern Out-of-Order (OoO) CPU, the components that are often subject of
%bottleneckness analyses are:
%\begin{itemize}
%	\item The \textbf{front-end}, which embeds the instruction decoding process and
%	produces sequences of micro-operations (\muop{s}) from the input program.
%	\item The \textbf{back-end}, or execution engine,
%	which embeds the computing units as well as 
%	the scheduling and dispatching components through which \muop{s}
%	access the formers.
%	\item The \textbf{speculative execution}, which pre-executes
%	instructions belonging to uncertain code paths (but degrades performance
%	in case of mispredictions).
%	%% which speculates on the program's
%	%% state to execute required instructions as soon as possible.
%	%% It improves program
%	%% performance when speculating correctly, but can significantly degrade it
%	%% otherwise in case of branch mispredictions or machine clears. 
%	\item The \textbf{memory subsystem}, corresponding to the cache hierarchy
%	and the off-chip memory.
%\end{itemize}}

\setlength{\jot}{8pt}
\begin{figure}[t]
	\footnotesize
	\begin{center}
		\begin{align*}
			\textit{Retiring} =\ & \frac{\textit{SlotsRetired}}{\textit{TotalSlots}} \\
			\textit{Front\ End\ Bound} =\ & \frac{\textit{FetchBubbles}}{\textit{TotalSlots}} \\
			\textit{Bad\ Spec} =\ & \frac{\textit{SlotsIssued} - \textit{SlotsRetired} + \textit{RecoveryBubbles}}{\textit{TotalSlots}} \\
			\textit{Back\ End\ Bound} =\ & 1 - (\textit{Front\ End\ Bound}\ +\ \textit{Bad\ Spec\ +\ Retiring}) \\
		\end{align*}
	\end{center}
	\vspace{-0.7cm}
	\caption{The four original formulas of TMA L1. On recent Intel microarchitectures, these are directly provided by ad-hoc PMC events.}
	\label{fig:topdown_formulas}
	\vspace{-0.5em}
\end{figure}

The TMA method has been implemented both in Intel's own \vtune{}~\cite{vtune} tool, as well as on the open-source \perf{}~\cite{perf} tool. 
Both use sampling to inspect PMCs that are used to classify pipeline slots into the categories mentioned above. 
Historically, the low-level counters shown in \figref{topdown_formulas} were used to calculate the relevant metrics, as described by A.~Yasin~\cite{topdown}. 
On recent generations of Intel CPUs, these metrics can be obtained through dedicated PMC events.

While TMA provides useful information about which parts of the microarchitecture act as bottlenecks for a given computation, it does not provide information about \emph{which parts of the code} trigger the bottlenecks, or why. 
Tools like \texttt{perf-annotate}~\cite{newperf} help correlate the obtained metrics to specific lines of code, but additional (human) analysis is required. 
Furthermore, this technique is subject to the inherent inaccuracy of sampling techniques~\cite{weaver2013non,pebs2020}.

In addition to sampling inaccuracies, TMA is also subject to methodological ones. 
Consider again the code in Fig.~\ref{lst:Motivation:Running-example}(a). 
When run on an Intel Core i9-12900k, the code achieves the aforementioned performance of 12.8~GFLOPS, and TMA indicates that the code is fundamentally B-E bound (87.5\%), and as memory-bound (44.1\%) as it is core-bound (43.4\%). 
Digging deeper into the TMA hierarchy, it finds that while the core is stressed by the utilization of the execution ports, the memory boundedness stems from pressure on the L1 cache. 
Given that on-peak FP computations are typically core-bound, the user, following TMA's guidance, decides to focus on the memory-bound component and optimize for register locality. 
The computation is rescheduled as shown in Fig.~\ref{lst:Motivation:Running-example}(b), on which the same \texttt{ymm2} value can be reused for the entirety of the inner loop.

After the optimization step, the code is re-run, but the same 12.8~GFLOPS performance is obtained. 
In this case, TMA indicates that the code is almost completely core-bound (91.4\%), due to the utilization of the execution ports in the B-E (71.4\%). 
In fact, both these computations are neither memory-bound nor bound by port capacity. 
The actual bottleneck lies on the dependence chain imposed by the reduction on \texttt{ymm0} performed by the \texttt{vfmadd} instructions, which limits the throughput to a single FMA at a time. 
The execution is therefore limited by the \emph{latency} of the FMA unit (4 cycles in this microarchitecture), yielding a throughput of a single \texttt{vfmadd}, i.e., 16 FLOPS, each 4 cycles at 3.2 GHz, that is, exactly 12.8~GFLOPS.

%, and in particular version (a), in which the entire inner loop computes directly on registers, without introducing any memory pressure. This code was run on a Golden Cove CPU (P-core from an i12900K Alder Lake processor) with AVX2. The TMA indicates that the code is fundamentally \emph{back-end bound} (92.5\%) and, more specifically, \emph{core bound} (91.4\%) and, going deeper, bound by the utilization of the execution ports in the backend (71.4\%). In fact, for the majority of the execution cycles (44.7\%), no \muop{} was launched for execution. This makes perfect sense: the \texttt{vfmadd} operations in the inner loop present a chained dependency, and therefore a single operation can be run at each cycle. The execution is therefore limited by the \emph{latency} of the FMA unit (4 cycles in this case), and presents a performance of precisely 4 FLOPs / cycle (one FMA, i.e., 2 FLOPs, on each of the 8 elements, issued each 4 cycles). The TMA appropriately characterizes the obtained performance, although it gives no hints as to the actual bottleneck of the computation: the dependence through \texttt{ymm0} which makes the code perform well below the theoretical peak of 48 FLOPs / cycle (which would be obtained with all 3x FMA units in simultaneous operation).

This TMA behavior stems from the fact that some bottlenecks cannot be \emph{causally} established by the usage of PMCs alone. Without any notion of dependencies among instructions and no knowledge of the code, TMA heuristically assumes that stalls caused by B-E pressure in the presence of an elevated Load Queue utilization must be counted towards the memory bound component, and is therefore misguided by this code.

\subsection{\gus's approach: Sensitivity and causality analysis}

The PMC-based analysis presented in the previous section relies on correlating high-pressured resources with stalls in order to identify performance bottlenecks. 
However, modern out-of-order architectures feature a high degree of instruction-level parallelism and complex resource interdependencies, and the fact that a resource is stalled while one it depends on is saturated does not necessarily mean that this latest has become a performance bottleneck.
%For instance, as presented in the previous section, the TMA method assumes that the code in Fig.~\ref{lst:Motivation:Running-example} becomes memory-bound by the fact that the load queue is under pressure during the entire computation, whereas in fact the latency of the FMA units is the fundamental driver of performance.

\begin{figure}
	\footnotesize
	\begin{center}
		\begin{tabular}{ l | c | c | c | c | }
			& p0 & p1 & p2 & p3\\
			\hline
			0 & $\mathtt{vfmadd}_{0,0}$ & & $\mathtt{vmovaps}_0$ & $\mathtt{vmovaps}_1$\\
			1 & && $\mathtt{vmovaps}_2$ &\\
			2 & && & $\mathtt{vmovaps}_3$\\
			3 & && $\mathtt{vmovaps}_4$ &\\
			4 && $\mathtt{vfmadd}_{0,1}$ & & $\mathtt{vmovaps}_5$\\
			5 & && $\mathtt{vmovaps}_6$ & $\mathtt{vmovaps}_7$ \\
			6 & && $\mathtt{vmovaps}_8$ & \\
			7 & && & $\mathtt{vmovaps}_9$\\
			\hline
			8 & $\mathtt{vfmadd}_{1,0}$ && $\mathtt{vmovaps}_{10}$ &\\
			9 && && $\mathtt{vmovaps}_{11}$\\
			10 && & $\mathtt{vmovaps}_{12}$ & $\mathtt{vmovaps}_{13}$\\
			11 && & $\mathtt{vmovaps}_{14}$ &\\
			12 && $\mathtt{vfmadd}_{1,1}$ && $\mathtt{vmovaps}_{15}$\\
			\vdots & \vdots & \vdots & \vdots & \vdots \\
		\end{tabular}
	\end{center}
	\caption{Port occupancy over time during the execution of version (b) of the inner loop from Fig.~\ref{lst:Motivation:Running-example}. Subindices indicate the iteration to which each instruction corresponds and, in the case of \texttt{vfmadd}s, whether it is the first or the second FMA in the loop.}
	\label{fig:timing}
\end{figure}

To illustrate this notion, \figref{timing} shows the port pressure during the execution of the first few cycles of the inner loop of the code in Fig~\ref{lst:Motivation:Running-example}(a). 
Ports p2 and p3, used for dispatching the \texttt{vmovaps} memory instructions, are under heavy pressure. 
TMA assumes correlation between this pressure with the stalls that lead to performance degradation. 
PMCs provide no way to realize that the dispatched \texttt{vmovaps} are out-of-order with respect to the \texttt{vfmadd}s. 
As such, TMA cannot discover that the \texttt{vfmadd}s are stalling due to dependencies with other \texttt{vfmadd}s, and not due to RAW hazards with the memory loads. 
The determination that any back-end stall in the presence of unresolved memory loads must be counted towards memory-boundedness is a heuristic one. 
However, as shown in \figref{timing}, it would be possible to introduce up to 7 additional memory loads in the inner loop before $p2$ and $p3$ became bottlenecks, decreasing the throughput from 8 to 9 cycles per iteration.\footnotemark

\footnotetext{Note that the difference in the number of \texttt{vmovaps} dispatched per cycle (one or two) depends on code alignment and the width of instruction cache ports.}

In this paper, a sensitivity-based approach is proposed. 
It consists in measuring the performance variations associated to the acceleration of different microarchitectural resources (e.g., port capacity, instruction latency, ROB size, etc.). 
This approach immediately reveals which factors directly affect performance. 
When applied to the codes in Fig.~\ref{lst:Motivation:Running-example}, the sensitivity-based analysis determines that the performance is linearly correlated with the instruction latency.

To find which static instruction is involved in the bottleneck, this paper also propose a causality-based approach.
It consists of propagating resources (and dynamic instructions) that constrain the availability (resp. retiring time) of resources (resp. instructions) that depend on them. 
In this specific example, it points to the chain of dependencies that constitutes the critical path of the code: 
$\mathtt{vfmadd}_{i,0} \rightarrow \mathtt{vfmadd}_{i,1} \rightarrow \ldots $. 
Those approaches are covered in detail in Sec.~\ref{sec:gus}. 
Identifying this bottleneck without sensitivity would require a detailed analysis of the execution, the code, and the dependencies between the different \muop{s}.

Being based on code instrumentation, \gus has yet another advantage over TMA: 
portability. 
Due to the very specific PMC events needed to implement the full hierarchy described by the TMA methodology, it needs a remarkably high level of support by vendors. 
As such, TMA is very well-supported on Intel's processors, but much more limited on AMD and ARM products, on which it is limited to a few hierarchical levels of detail.

\subsection{Other performance analysis techniques}

Although Intel's TMA has become one of the most popular performance analysis techniques, there are other approaches that are covered in the following paragraphs.

\subsubsection{Static binary analysis}
All the tools described here take a basic-block in isolation as an input and statically analyze it as if it were the body of a loop.

Intel \iaca~\cite{iaca} was the code analyzer distributed by Intel until 2019. 
When applied to a basic block such as the one in Fig.~\ref{lst:Motivation:Running-example}, it delivers a report including a throughput estimation, a detailed ports pressure analysis and a bottleneck analysis which outputs the most probable bottleneck cause.
In \iaca, bottleneck analysis essentially consists in looking for the source of probable \textit{stalls}. 
It first looks for a traffic jam in the back-end of the CPU and, if it finds one, considers it the bottleneck. 
Otherwise, it computes the throughput of the basic block when ignoring latencies. 
If a significant difference with the nominal throughput prediction is found, the basic block is latency-bound. 
Finally, if the bottleneck is neither the back-end nor instruction latency, then the front-end is marked as bottleneck. 
The \llvmmca~\cite{llvmmca} code analyzer is based on a similar approach. 
Its bottleneck analysis engine exploits the underlying performance model used for throughput estimation. 
It provides indicators of back-end resource saturation and of the influence of dependence-induced latencies on performance.

\facile and \uica use automatically generated performance models, and differ simply in the algorithm used to make the basic block throughput estimation, which influences the bottleneck analysis. 
On the one hand, \uica~\cite{uica} associates the throughput of a basic block with the timestamp at which its last instruction was retired (after possibly many iterations). 
A resource is considered to be a bottleneck if its throughput exceeds a certain hard-coded proportion of the total system throughput. 
On the other hand, \facile~\cite{facile} considers that the basic block throughput is the throughput of the slowest CPU resource. 
With this approach, the latter is by definition the (only) performance-limiting bottleneck.

Both Intel~\iaca, \llvmmca, \facile, and \uica are meant to provide bottleneck analysis.
Other tools in this category are limited to throughput estimation. 
It is the case of neural-network based ones such as \granite{}~\cite{granite} and \ithemal{}~\cite{ithemal}. 
Others only partially model the microarchitecture. 
For instance, \osaca{}~\cite{osaca} assumes that a basic block is never front-end bound.

All the static tools listed above are known for suffering from a lack of
context~\cite{anica}:
by ignoring actual values of variables when entering the loop, and by not simulating the evolution of scalar variables they are enabling to model the flow of data through cache and memory leading to critical inaccuracies, even for L1-resident computing kernels.

\subsubsection{Simulation-based analysis}

A possible approach to estimating performance is the use of simulators which give insight into what happens inside the CPU during the execution of an application, without the need to actually measure events in real hardware. 
This allows to estimate the performance of full or isolated subsystems without actually building them, and for this reason it is the predominant approach in microarchitectural exploration works. 
gem5~\cite{gem5} is an architectural simulator capable of simulating out-of-order execution. 
More details about how gem5 relates to \gus are given in Section~\ref{sec:gus}.
In general, the use of cycle-level simulators for performance debugging is not a popular approach, due to two fundamental limitations. 
The first one is methodological: 
building a cycle-level model of all the events taking place inside a CPU is a very different problem from actually understanding its performance. 
Similar to the limitations described for TMA, it is necessary to build causal models that relate instruction latency, port capacity, dependencies, and code. 
The second one is accuracy: 
architectural simulators are useful tools for differential analysis, i.e., discovering the relative differences between two different designs, but they are not particularly accurate at predicting the actual execution time of a kernel on a given hardware. 
To overcome these limitations, the current work turns to binary code instrumentation, on which a coarse-grained architecture is simulated in order to discover first-order performance effects. 
This approach offers a much more accurate performance prediction, as detailed in Sec.~\ref{sec:experiments}.

%% file: src/running-example.tex
\begin{figure}
  \begin{subfigure}[t]{0.25\textwidth}
    \begin{lstlisting}[language={[x86masm]Assembler},escapechar=@,basicstyle=\tiny]
vmovaps (%rcx),%ymm1            
for
  vmovaps (%rax),%ymm2             
  vfmadd231ps %ymm0,%ymm1,%ymm3   
  vfmadd231ps %ymm0,%ymm1,%ymm2   
\end{lstlisting}
    \caption{}
  \end{subfigure}%
  \begin{subfigure}[t]{0.25\textwidth}
    \begin{lstlisting}[language={[x86masm]Assembler},escapechar=@,basicstyle=\tiny]
vmovaps (%rcx),%ymm1
vmovaps (%rax),%ymm2             
for
  vfmadd231ps %ymm0,%ymm1,%ymm3   
  vfmadd231ps %ymm0,%ymm1,%ymm2   
\end{lstlisting}
     \caption{}
  \end{subfigure}

  \caption{\label{lst:Motivation:Running-example}
    An example in pseudo-asm code of a kernel computing \texttt{ymm0 = ymm1*ymm3 + ymm1*ymm2} iteratively. Integer operations and branches supporting pointer arithmetic and loop iteration have been removed for clarity. \texttt{ymm1} and \texttt{ymm2} are loaded from two non-overlapping memory arrays. \texttt{ymm3} is constant. In version (b), the \texttt{vmovaps} to \texttt{ymm2} is hoisted out of the inner loop.}
  
\end{figure}

%% file: src/03-gus.tex
%-------------------------------------------------------------------------------
\section{\gus, a sensitivity-oriented code analyzer}\label{sec:gus}
%-------------------------------------------------------------------------------

We introduce in the following \toolsensitivity{}, a microarchitectural simulator for performance debugging.
The design of \toolsensitivity{} is based around three fundamental goals:

\paragraph{Portability}

The design of \toolsensitivity decouples the functional part of the execution from the architectural aspects that drive performance.
In the front-end of the tool, an emulator (QEMU) captures the functional behavior of the binary to be analyzed.
It then uses binary instrumentation to feed an abstract simulator that models the performance of and interactions between the different architectural blocks.
Decoupling the functional and architectural aspects allows using abstract performance models, including those developed for compilers or analyzers such as \llvmmca or \uica.
This enables the integration of performance models generated automatically through micro-benchmarking, ensuring portability, which is crucial for precision.
These performance models can be improved orthogonally to the development of the tool, benefitting precision and simulation performance.

This design contrasts with traditional architectural simulators such as Gem5, which integrate both aspects and simulate executions by modeling the behavior of each architectural block in an extremely detailed way. This immensely complicates the implementation of new architectures, often leading to ``simulation by proxy'' approaches which introduce inaccuracies.

%Using a simulator instead of hardware support to monitor the host architecture's behavior during program execution offers the flexibility to implement any complex instrumentation mechanism. 
%However, this approach significantly slows down execution compared to native performance, so addressing speed is crucial. 
%When designing a simulator, accurately modeling the actual behavior is essential. 
%Specifically, using a simulator that models one micro-architecture's behavior to another's introduces inaccuracies. 
%\toolsensitivity{} is based on a simulator that separates functional behavior from behavioral aspects. 
%Unlike Gem5, which integrates these aspects to simulate a program's execution on a given architecture by detailing each component's behavior and interactions (including the front-end decoder), our design uses an emulator (QEMU) to capture functional behavior. 
%We then use binary instrumentation to feed an abstract simulator that models interactions between components affecting performance. 
%This approach allows for a best-effort performance model. 
%Moreover, we can utilize abstract performance models developed for compilers like LLVM-MCA. 
%This community's approach involves automatically generating performance models using micro-benchmarks, ensuring portability which is crucial for precision. 
%Improving these models is ongoing and is separate from this work's contribution. 
%This decoupling enhances both portability (and thus precision) and speed (reducing slowdown).
\paragraph{Sensitivity}
As previously mentioned, focusing solely on resource usage rates is insufficient on its own, and can lead to incorrect bottleneck characterization. 
%Also the enhancement provided by TMA, which refines resource usage monitoring (e.g., by measuring when a resource is used while another is available), is insufficient on its own. 
Additionally, bottlenecks like latency issues are hard to detect through resource usage alone. 
In \toolsensitivity{}, the key criterion for identifying a resource as a bottleneck is sensitivity analysis: 
adjusting its capacity (throughput or latency) to verify its effect on the performance of the system.
%if performance degrades (or improves). 

An example of sensitivity-based analysis can be borrowed from the HPC realm, on which users often characterize the bottleneck of a computation by lowering the frequency of the CPU through DVFS while keeping the frequency of the memory constant.
%This is similar to what is done in HPC by lowering the CPU frequency while keeping the memory frequency constant through DVFS. 
If this does not slow down the kernel's overall execution, it is indicative of memory boundedness.
Generalizing this approach using a simulator is another feature of \toolsensitivity{}. 
To enable such analysis, the simulator must allow for parameterizing resource capacity. 
Unlike the detailed behavioral simulation in Gem5, our abstract model-based design is particularly suited for this purpose.
\emph{Given this ability of changing the capacity of each individual resource in our simulator, the implementation of sensitivity in \toolsensitivity{} is pretty simple:
for each abstract resource, the simulator is run with increased capacity (e.g., decreasing the instruction latencies to pinpoint latency-bound kernels, or increasing the L2 cache bandwidth to pinpoint L2-bound kernels) and the predicted execution time is compared.} 

\paragraph{Causality}
While identifying the set of instructions that stress execution ports is straightforward, it is more challenging to do the same for other bottlenecks like the ROB or instruction latencies in ILP-bound kernels. 
The analysis we developed in \gus to highlight bottleneck instructions is called \emph{causality analysis}. 
Its basic idea is to use a ``tainting'' mechanism during simulation to trace the instructions or resources that constrain the execution time of others. 
It is important to note that such propagation would not be possible with an event-based simulator e.g.\ relying on a SystemC~\cite{systemC} architectural description. 
Instead of this standard technique, we designed our simulator based on constraints propagation. 
This technique does not provide the state of each resource at any given time. 
Instead, it processes the stream of decoded instructions and maintains the earliest (optimistic) availability time for each resource (including hardware resources, operands, instructions, etc.). 
When a resource is needed, it checks the resources it depends on and propagates time constraints (usually taking the maximum).
\emph{This approach is enabled by using conjunctive resource mapping, as described in~\cite{palmed}, rather than the standard port mapping used in~\cite{uops,pmevo}.
This property of the simulator allowed us to develop our tainting-based causality analysis, which we will describe below.}

The remainder of this section covers the design and implementation details of \gus. 
%A brief overview of its global design is given in \figref{overview}.
%First, we explain the design choices that led to the creation of an efficient  instruction driven simulator (\ding{202}). 
First, we detail the underlying performance model of \gus based on abstract resources. 
Then, we explain how sensitivity-based analysis is used in \gus.
Finally, we demonstrate the usefulness of sensitivity and how it overcomes the shortcomings of previous bottlenecks detection techniques.

%\begin{figure}
%    \begin{center}
%      \includegraphics[width=\linewidth]{figures/overview.pdf}
%    \end{center}
%    \caption{Overview of \toolsensitivity{}'s architecture.}
%    \label{fig:overview}
%    \vspace{-1em}
%\end{figure}

%\subsection{A dynamic binary instrumentation based frontend}
%
%The front-end of \toolsensitivity{} is built around \qemu~\cite{qemu}, a fast
%functional processor emulator. It relies on user-mode dynamic binary
%instrumentation within a \qemu{}~plugin to generate the stream of events that
%drive the simulation.

% Not very useful
% This is mostly motivated by the fact that execution traces can become
% extremely large, thus storing them on disk is not usually an option for
% scalability reasons.
% Consequently, while static code analyzers~\cite{uica,iaca} typically work on a
% single basic block level, \toolsensitivity{} can work
% a much larger scope chosen by the user, for instance a function call
% or a whole program.

% Furthermore, a trace gives \toolsensitivity{} a lot of
% information about the execution of the program, such as whether a branch was
% taken or not, if two memory accesses alias, \etc{} By comparison,
% static code analyzers have to make assumptions that
% may not hold in practice. Memory dependencies are a good example of this, as
% most code analyzers ignore them completely, that leads to imprecise
% results~\cite{anica}.

\subsection{An abstract resource-centric performance model and causality analyser} 

\input{src/algo-gus-tainting.tex}

Algorithm~\ref{alg:gus} shows in detail the coarse-grained simulation approach implemented by \toolsensitivity. 
We first ignore tainting.
The algorithm takes as input a stream (T) of instructions provided by QEMU.
For the current instruction \texttt{i}, it updates the state of resources used during its different execution stages.
Before \texttt{i} can be dispatched, the dispatch queue (IDQ) must have an empty slot.
The earliest time at which this can be done is represented as \dispatch.\tavail. 
This value is updated in line 21 by constraining it to be greater than the time at which the latest instruction in the IDQ can be popped, that is, its retiring time.
The resource itself that decodes and dispatches the instruction is represented in this pseudocode as \frontend.
Here we assume it has an inverse throughput of \frontend.\ithr.
The earliest time at which this abstract resource can be used is represented as \frontend.\tavail.
Just as for any abstract resource, in our simulator this value can only increase.
When increased by its inverse throughput in line 16 it means it is used during a time \ithr.
When increased (line 13) because \tmin is greater than \frontend.\tavail (line 11) it means it is idle between \frontend.\tavail and \tmin.
It must be noted that this does not reflect actual usage but provides an approximation of a lower bound on the available timestamp of a resource.

Only once \texttt{i} has been decoded can it be dispatched.
This constraint is expressed on line 24.
Instruction \texttt{i} is pushed on the dispatch queue and the earliest time at which it can be executed is set to this timestamp.
We continue to follow the lifetime of \texttt{i} by traversing all the locations of its memory operands.
A simple cache simulator 
%(here DineroV~\cite{dinerov}, with pseudo-LRU replacement policy) 
tells at which cache level the corresponding cache line \texttt{loc} of the operand data is located.
$\Ll$ (with $i$ the cache level) is an abstract resource that models the throughput between $\Ll$ and the previous cache level.
Line 29 states that earliest time at which \texttt{loc} can be made available in L1 is not lesser than the earliest availability of $\Ll$.
The limited bandwidth between $\Ll$ and the previous cache level is represented as $\Ll.\ithr$.
Similarly to any other abstract resource, its usage is modeled by incrementing $\Ll.\tavail$ by $\Ll.\ithr$.
Also, a stall in the dispatch queue will create a bubble in this resource equal to $\tmin-\Ll.\tavail$.

Once the cache related resource states are updated, the constraints imposed by read operands on the time at which \texttt{i} can be executed are propagated to \texttt{i.\tavail}.
A subtlety is that here \texttt{loc} is a word while before it was a line.
The shadow memory which is used to store this information contains both information at the granularity of bytes and information at the granularity of lines.
The one at the granularity of bytes is both constrained by the cache behavior, but also by the earliest time at which the last instruction that modified it could be executed.

Similarly to previously considered abstract resources, the \texttt{\tavail} of the resources \texttt{res} that represent the compute units (combined ports) are taken into account to update \texttt{i.\tavail} (line 34), and inversally the \texttt{res.\tavail} is increased accordingly (line 35).
The rest of the code reflects the next stages where \texttt{i.latency} is the expected execution latency (obtained using microbenchmarks) of \texttt{i}.

The tainting does nothing else than propagating a set of instructions from the constraining resource (\texttt{c} in line 1) to the constrained instruction (lines 32 and 34) or resource (lines 21, 22, 24, 29):
if the constraining resource \texttt{c} creates a bubble (line 4), then the tainting of the constrained resource (this in line 1) is cleared and set to the tainting set of \texttt{c} (line 6).
Here, the semantic of \texttt{this.taint} corresponds to the set of all dynamic instructions that may impact the \texttt{\tavail} of this one.
In case of equality (line 2) both the original set of \texttt{this.taint} and the set of \texttt{c.taint} are possible contributors (line 3).
Obviously the tainting does not keep all past dynamic instructions.
Instead, a queue is used.
The pop of \texttt{i'} in line 42 starts only once \texttt{taintqueue} is sufficiently large (its length is set to twice the ROB size by default).
As such, \texttt{i'} corresponds to a sufficiently old instruction.
For this instruction, we consider it to contribute to the overall execution time of the program if it delayed the dispatch of \texttt{i}, which is reflected as soon as it belongs to the tainting set \texttt{dispatch.taint}.

\subsubsection{Implementation details}
\label{sec:resources}
%
%The building blocks of the model are the abstract throughput-limited resources.
%In order to execute an instruction, one or more of these resources may be used
%and must be available. Otherwise, the instruction is stalled until all the
%required resources become available. Each resource tracks the time at which it
%will be available to accept another request, this is characterized by its
%throughput.
%
%The timestamp at which a resource \(R\) is available to accept an instruction is
%denoted as \(R.\tavail\). Every time a resource is used, its \(\tavail\) is
%incremented by its inverse throughput (or \emph{gap})
%(\cref{alg:cacheavail,alg:tavail}), which determines the minimum amount of
%time that must elapse between two uses of the resource.

\paragraph{Front-end}

The front-end is modeled by two resources representing 1) decoding unit whose behavior is described by \texttt{uiCA} \cite{uica}
and 2) the \muop{} cache tied to a resource of fixed throughput.
%The F-E is modeled as a single resource with a throughput of $I_w$
%instructions per cycle, where $I_w$ corresponds to the issue width of the simulated architecture.
%This assumption may hide some fine-grained bottlenecks.

\paragraph{Execution ports and functional units}

The classical formalism to describe the throughput and sharing of back-end
resources is a port mapping, a tripartite graph which describes how
instructions decompose into \muop{s} and which functional units \muop{s} can
execute on.

\toolsensitivity{} uses a simpler two-level representation~\cite{palmed} instead, called a resource mapping, where instructions are directly associated with a list of abstract resources. 
To account for \muop{} decomposition, a resource can appear in this list multiple times. 
Aside from avoiding the need to explicitly model the port scheduling algorithm in the back-end and thus allows to achieve a lower runtime overhead, this representation allows a simple propagation of constraints which constitutes the foundation of causality analysis.

\paragraph{Latencies}

The latencies used for execution units in our experiments are extracted from
\uopsinfo~\cite{Abel19a} and used to build a resource mapping. For bottleneck
analysis, we present latencies induced by dependencies as a separate resource.
The same approach can be found, for instance, in \iaca{}. The latter helps guide
optimization by highlighting the need to discover more parallelism, rather than
the need to relieve the load on a hardware component.

\paragraph{Dispatch Queue}

The ROB capacity is modeled by the use of a
dispatch queue, i.e., a finite-sized buffer of instructions that are
in-flight (issued but not retired).

The dispatch queue is bounded by the number of slots it has. As such, if the
queue becomes full, no more instructions can be issued until one
is retired. 
%We denote \tmin{} as the earliest time a slot in the window will become free (\cref{alg:tmin}).

%\paragraph{Shadow memory and register file}
%
%The shadow memory and shadow register are used to detect data dependencies
%between instructions. 
%For each memory cell or register they store the time at which the data in this
%location will be available. 
%The update mechanism is twofold:  (i) when an instruction writes to a location, the shadow cell for that
%location will be set to the time when the instruction is retired
%(\cref{alg:shadowvalue}); 
%and (ii) when a cache miss occurs this counts as a use
%of all levels of the memory hierarchy up to the one where the miss occurred. The
%location in the shadow memory corresponding to the accessed memory location is
%then updated to the maximum availability of the involved resources
%(\cref{alg:cacheavailread}).

\paragraph{Caches}

\toolsensitivity{} uses a fork~\cite{forkdinero} of the Dinero~IV~\cite{dineroiv} simulator to determine hits and misses in the different
levels of the cache hierarchy, enriched with access latencies and a next-line prefetch heuristic.
The PLRU (Pseudo-LRU) replacement policy from
this fork is used for all levels of the hierarchy. However, cache
replacement policies implemented in commercial processors are a complex
topic~\cite{Abel20}, and an exhaustive modelization is out of the scope of
\toolsensitivity{}.
% that acts mostly as a generic modern out-of-order CPU model.

% Not in mainstream GUS
% \paragraph{Branch prediction unit simulator}

% The branch prediction unit is composed of a branch target buffer (BTB) modeled
% as an associative array with an LRU replacement policy, and accompanied by an
% indirect branch predictor (IBP) with a ITTAGE scheme~\cite{Seznec2011ITTAGE} and
% a conditional branch predictor (CBP) with an L-TAGE
% scheme~\cite{Seznec2007LTAGE}. Findings in the
% literature~\cite{rohou2015ibp,yavarzadeh:2023:half, zen2} show that such schemes
% or close derivatives are actually used in \texttt{x86} commercial processors and
% thus good candidates for the model of \toolsensitivity.

%% \subsubsection{Core performance model}

%% To reduce simulation times, some simplifying assumptions are made:

%% \begin{itemize}
%%     \item \toolsensitivity{} assumes that the program has regular access memory
%%     patterns with latencies that can be perfectly hidden by prefetching,
%%     similarly to what is done in the \ecm{}~\cite{hager2016, johannes2018} model.
%%     \item \toolsensitivity{} does not model neither load/store queues nor
%%     load-store forwarding. As such, the bandwidth between the CPU and L1
%%     is considered infinite and not modeled as a resource.
%%     \item \toolsensitivity{} also does not model execution pipeline hazards or
%%     operand forwarding.
%% \end{itemize}

%% \noindent{}This high-level model is sufficient to achieve state-of-the-art precision as
%% we will see in Sec.~\ref{sec:experiments}.

\subsection{A sensitivity-oriented code analyzer} 

Sensitivity analysis allows pinpointing the source of a performance bottleneck precisely across instructions. 
We briefly describe here how it compares with other related works and how it is implemented and used in \toolsensitivity{}.

Sensitivity analysis~\cite{hong_sensitivity_2018,kolia_2013}, also called
differential profiling~\cite{mckenney1995}, works by executing a program multiple
times, each time varying the usage or capacity of one or more resources.
Bottlenecks are then identified by observing how these changes impact the overall performance, \ie{} how sensitive performance is to each resource.

Moreover, sensitivity analysis is an automatic approach that does not require as
much expertise as existing performance debugging
tools~\cite{cqa_2014,kerncraft2017}. 
This approach requires careful consideration of different trade-offs in order to be practical. 
%Modern hardware does not offer many ways to easily vary the
%capacity of resources, motivating the need for a model such as the one in
%\toolsensitivity.

\paragraph{Simulation granularity}

\toolsensitivity employs a coarse-grained simulation approach known as
'Instruction-window centric simulation'~\cite{genbrugge2010, carlson2014}. 
The fundamental concept of this simulator category~\cite{sanchez_zsim_2013,sniper2011} 
is to represent only components that are critical 
for performance, thereby enhancing simulation speed.
As such \toolsensitivity's model assigns a fixed, per-instruction cost to each simulated resource, 
driven by reverse-engineered microarchitectural features.

This is computationally less intensive than the usual approach 
in cycle-level simulators such as gem5~\cite{gem5}.
The latter embeds a single model to simulate OoO execution, the \texttt{O3CPU} model. 
While this model is designed to be configurable to accurately reflect the microarchitectural layout, 
some components \emph{must} be defined for it to be usable: 
fetch and decode logic, structures for issuing instructions such as the issue queue, 
register renaming, or pipeline stages.

Some modes (FSA~\cite{sandberg_2015} or functional mode) are designed to \emph{swiftly} establish 
a consistent checkpoint of Gem5's internal structures  at a predetermined point in the execution process (usually post-boot), 
foregoing any simulation. 
A switch to a detailed model, based on the \texttt{O3CPU}, is then done for any kind of simulation.
Similarly, \toolsensitivity can also restrict its instrumentation to a subset of the program, leaving the other code 
generated by \qemu{} to run as is.
While this reduces the computational cost of the simulation by a significant margin, the bottleneck of running the region of interest itself still remains. The coarse-grained approach implemented by \toolsensitivity{} has proved to be a good compromise between  simulation speed and precision.

\paragraph{Sensitivity analysis in the broader world}

The idea of varying microarchitectural features is not new in itself either, as
it has been used by hardware designers for Microarchitecture Design Space
Exploration~\cite{ipek_asplos_2006, bai_archexplorer_2023} to guide
microarchitecture parameter tuning to explore the trade-offs amongst
performance, power, and area (PPA).

\decan~\cite{kolia_2013} is a dynamic performance analysis tool based on the
\maqao~\cite{DBCLAJ05a} binary analysis and instrumentation framework. \decan{}
finds bottlenecks by sensitivity analysis based on binary rewriting; it
removes or modifies instructions in a kernel and checks by how much
each transformation affects the overall performance via performance counters.

\coz~\cite{coz} relies on \textit{causal profiling}, which consists in
sensitivity on code lines. Each one is successively ``accelerated'' to quantify its
impact on performance, by slowing down the surrounding code (in practice, the latter
is embedded in threads that are paused when necessary).

% The low overhead of this approach allows it to quickly explore a large set of variants for its
% sensitivity analysis. The downside of \decan{}’s approach is that its
% transformations are of course not semantic preserving and can easily introduce
% crashes or floating-point exceptions. Changing the semantics of a program like
% this might, of course, also change its performance behavior in other subtle
% ways, making it hard to verify or falsify the results produced by the tool.

The \saake{} system~\cite{hong_sensitivity_2018} is conceptually closer to
\toolsensitivity{} in its implementation of sensitivity analysis. It uses a fast
symbolic execution engine that estimates the runtime of GPU programs to drive
sensitivity analysis for finding bottlenecks.

% \saake{} input independent
% abstract simulation works well for the simpler microarchitectures of GPUs since
% they do not use out-of-order execution or speculation and handle branching
% control flow using predicated execution. Since \saake{} does not actually
% simulate the execution of instructions, there are several things it can not
% compute that have to be provided externally.

\paragraph{Sensitivity analysis in \toolsensitivity}

%% fp(wr) <= fp(r)
%% hence formula is:
%% speedup = fp(r)/fp(wr) - 1

%% Actually better to name resource r, thoughput of resource Tr and thus accelerated thoughput: w.Tr
%% Hence:
%% Sw,r = Fp(Tr)/Fp(w.Tr) - 1

In addition to execution ports, resources which are subject to sensitivity
analysis in \toolsensitivity are: the instruction latencies, the
front-end throughput, the size of the ROB, the throughput of the retire buffer,
and the communication bandwidths between the different memory levels.

Each resource has a \textit{capacity}, a finite quantity whose variation
may impact the execution time of a program.
These may be quantities immediately linked
to a hardware component, such as the size of the ROB, or more abstract
in nature, such as instruction latency.
We represent each of these quantities by a real number.
The sensitivity analysis implemented in \toolsensitivity{} consists in
varying the $n$ resources represented in the model during
successive execution time estimates performed on the same program $p$.
At each iteration, the capacity $c_r$ of a resource $r$ is successively
weighted by real numbers $w_0,...,w_m$ to discover a $w$
that minimizes the estimation function $f_p$. In this framework, the other
resource capacities and the input program $p$ can be seen as constants.
Thus, $f_p$ is a function from real numbers to real numbers, 
associating an estimate of duration with $c_r$.
The speed-up $s_r$ obtained by weighting $c_r$ by $w$ is therefore calculated
as follows:
\begin{equation*}
s_{w,r} = \frac{f_p(c_r)}{f_p(w*c_r)} - 1
\end{equation*}

Resources whose variations result in a speedup are
bottleneck resources, and
should be the focus of optimization efforts.

%% In the general case, the complexity of this algorithm is $O(n*m)$, $n$
%% being the number of the model resources, and $m$ the number of weight
%% candidates applied to each of the latter.
%% However, at a first glance, looking at a single weight value
%% may be sufficient for answering where are the bottlenecks of a program before
%% precisely quantifying their overall impact.

\paragraph{\toolsensitivity instruction-level report}

%% We use a form of heat-map for the
%% visualization. Each bar in it represents one abstract resource. The height of
%% each bar and its color indicate the speedup predicted by \toolsensitivity{} if
%% the throughput of that resource is increased.

%% \begin{figure}
%%     \begin{center}
%%       \includesvg[inkscapelatex=false, width=\linewidth]{figures/sensitivity.svg}
%%     \end{center}
%%     \caption{Sensitivity analysis report produced by \toolsensitivity{}
%%       (Skylake architecture) for the
%%       Jacobi iteration method implementation discussed in
%%       \secref{intro}.}
%%     \label{fig:sensitivity}
%%     \vspace{-1em}
%% \end{figure}

%% \figref{sensitivity} depicts the view of bottlenecks for the Jacobi example
%% discussed in \secref{intro}. According to \toolsensitivity, the runtime of the kernel
%% is not affected by increasing the throughput of any other resource aside than ports 2 and 3. 
%% Hence, \texttt{p23} is the sole bottlenecked resource.

%% In addition to execution ports, resources which are subject to sensitivity
%% analysis in Gus are: the instruction latencies (\texttt{INST LAT}), the
%% front-end throughput (\texttt{FRONTEND}), the size of the ROB (\texttt{INST
%% WINDOW}), and the communication bandwiths between the different memory levels
%% (\texttt{L2 THR}, \texttt{L3 THR} and \texttt{MEM THR}).

\input{src/gus-raw-report.tex}

One can also look at the resource usage per instruction to pinpoint it in the
kernel. Table~\ref{tab:raw-report} contains the instruction-level report generated by \toolsensitivity from a Jacobi stencil. Sensitivity analysis shows
that the runtime of this kernel is not affected by increasing the throughput of any
other resource aside than ports p2 and p3, but the report is more specific.
It shows that multiple instructions (and which) use the resource \texttt{p23}
by loading from memory.
%% This is done by looking at the fine grain report shown in \cref{tab:raw-report}.
%% We see that multiple instructions use the resource \texttt{p23} by loading from
%% memory. Reducing this through a transformation (e.g: register tiling) would reduce the 
%% pressure on this resource and could lead to a speedup of the kernel.
This level of precision is useful for performance
debugging, as it allows a level of granularity that is not
usually possible through the measurement of PMC events~\cite{topdown}.

\subsection{Using \toolsensitivity for manual code optimization}

In this subsection, we illustrate how sensitivity analysis can be used to drive manual code optimization.
For this purpose we analyze the \texttt{correlation} kernel,
which is one of the evaluation benchmarks discussed below
(in \autoref{ssec:framework}).
Its most costly loops nest is shown in Fig.~\ref{fig:correlation}.
The different steps are reported in Table~\ref{tab:gus-loop}.
The compiler (gcc-13 in this example) does a very poor job of vectorizing the original version, leading to a very low performance.
We thus start by vectorizing the code by hand, leading to version \textbf{v1}, which is still not very efficient.
On this version, both TMA and \toolsensitivity{} pinpoint L2 bandwidth as a bottleneck.
We apply a register tiling transformation to improve data locality.
Considering the target microarchitecture (Skylake-X),
we choose a 2x16 register tile size
With AVX2, this executes 4x SIMD Fuse-Multiply-Adds (FMAs) per iteration, 2 iterations running in parallel.
This version \textbf{v2} reaches 33.1\% of the peak performance.

On this new version, \toolsensitivity{} highlights the retire buffer as the bottleneck. The
detailed report shows that the stores (\texttt{vmovapd}) cause this behavior as they take 2 slots in the retire buffer.
We hoist the load/store instructions outside of the innermost loop to reduce the overall number of instructions, thus the
pressure on the retire buffer.
The obtained version \textbf{v3} reaches 46.8\,\% of the peak performance.
\toolsensitivity{} pinpoints now the L2 cache accesses
as the bottleneck. We tile the axis of the matrix
multiplication in order to improve L1-reuse.
This optimization (leading to version \textbf{v4}) unlocks 81.2\,\% of the peak performance.
\toolsensitivity{} pinpoints the retire buffer by showing too many \texttt{vmovapd} suggesting a more squared shaped register tile.
Forgetting that our architecture contained only 16 vector registers, we had inadvertently changed the register tile to 4x16.
Sensitivity analysis on \textbf{v5} outputs instruction latency as a bottleneck, introduced by dependency chains on spilled registers; which
lead to our final version \textbf{v6}.
There, the bottleneck shifts to ports p0 and p1 that process FMAs. As these are exactly the core operations we
want to execute, we stop here the optimization process.

Overall we improve the performance by 662.4\,\%, achieving 82.8\,\% of the peak performance. 
As summarized in Table~\ref{tab:gus-loop}, optimization is guided at every stage by \toolsensitivity{}'s sensitivity analysis and instruction-level report. 
TMA provides little guidance from step 3 onwards.
In particular in step 5, while TMA points to backend-boundedness, which is not straightforward to detect spilling:
only sensitivity analysis on instruction latency assures that dependency chains are the bottleneck.

\begin{figure}[!htb]
\vspace{-1em}
\begin{lstlisting}[language=C,escapechar=|]
for (int j = 0; j < _PB_M; j++) {
  for (int i = j; i < _PB_M; i++) {
    for (int k = 0; k < _PB_N; k++) {
      corr[i][j] += data[k][i] * data[k][j];
    }
    corr[j][i] = corr[i][j]; |\label{line:correlation:transpose}|
  }
}
\end{lstlisting}
\caption{Untransformed correlation kernel.}
\vspace{-1em}
\label{fig:correlation}
\end{figure}

\begin{table*}
  %% \resizebox{\columnwidth}{!}{
  \footnotesize
    \begin{tabular}{l l l l l }
      \toprule
      Version & Transformations & Peak perf. & TAM (perf) & Bottleneck (\toolsensitivity{}): ressource, instructions\\
      \midrule
      v0 & None & 4.5\% &               &                  \\
      \midrule
      v1 & \emph{Vectorized} & 12.5\% & Backend(L2) & L2\\
      \midrule
        v2 & Vectorized, \emph{Register tiling 2x16} & 33.1\% & Retiring & Retire Buffer \\
      \midrule
      v3 & Vectorized, Register tiling 2x16, \emph{Hoisting} & 46.8\% & Backend (ports usage) & L2, \texttt{vbroadcastsd}, \texttt{vmovaddpd} (loads)\\
      \midrule
      v4 & Vectorized, Register tiling 2x16, Hoisting, \emph{Tiling} & 81.2\% & Backend (ports usage) & Retire buffer, includ. \texttt{vmovaddpd}\\
      v5 & Vectorized, Register tiling \emph{4x16}, Hoisting, {Tiling} & 52.6\% & Backend (ports usage) & Latency, includ. \texttt{vmovaddps[rsp]} (spill)\\
      v6 & Vectorized, Register tiling \emph{3x12}, Hoisting, {Tiling} & 82.8\% & Backend (ports usage) & P01\\
      \bottomrule
      
    \end{tabular}
    %% }
    \vspace{1em}
    \caption{Optimization process  using \toolsensitivity{} on the
      most costly loops extracted from the benchmark \texttt{correlation}.}
  \label{tab:gus-loop}
  \vspace{-1em}
\end{table*}

% sensitivité / instructions en jeu / ressources fortement utilisées/saturées / Transfor / Perf
% transpose: 3% du jeu à discuter
% highlighter instructions assembleurs

%% file: src/algo-gus-tainting.tex
\begin{algorithm}
    \scriptsize 
    \caption{Core algorithm of \toolsensitivity{}}
    \label{alg:gus}

\Function{ConstrainBy(\this, \c)}{
    \If{$\this.\tavail=\c.\tavail$}{
        $\this.\taint \assign \this.\taint \cup \c.\taint$\;
     }\ElseIf{$\this.\tavail<\c.\tavail$}{
         $\this.\tavail \assign \c.\tavail$\;
         $\this.\taint \assign \c.\taint$\;
     }
}

\Function{SetBy(\this, \c)}{
     $\this.\tavail \assign \c.\tavail$\;
     $\this.\taint \assign \c.\taint$\;
}

\Function{UsedBy($\this, \i, \tmin$)}{
  \eIf{$\this.\tavail<\tmin$}{
      $\this.\taint \assign \{\i\}$\;
      $\this.\tavail \assign \tmin$
  }{$\this.\taint \assign \this.\taint \cup \{\i\}$}
  $\this.\tavail \assign \this.\tavail + \this.\ithr$\;
}

\Function{Simulate()}{
 \ForEach{$\i \in \T$}{
   $\taintqueue.\var{push}(\i)$\;
   \Comment{IDQ/retiring}
   $\i_\var{retired} \assign \dispatchqueue.\var{pop}()$\;
   $\dispatch.\FuncCall{ConstrainBy}{$\i_\var{retired}$}$\;
   \Comment{Front-end}
   $\frontend.\FuncCall{ConstrainBy}{\dispatch}$\;
   \frontend.\FuncCall{UsedBy}{$\i, \tmin=\dispatch.\tavail$}\;
   \Comment{IDQ/dispatch}
   $\dispatch.\FuncCall{ConstrainBy}{\frontend}$\;
   $\dispatchqueue.\var{push}(\i)$\;
   $\i.\FuncCall{SetBy}{\dispatch}$\;
   \Comment{Cache}
   \ForEach{memory line \loc accessed by \i}{
      \ForEach{cache level $\Ll$ hit by access of \loc}{
         $\loc.\FuncCall{ConstrainBy}{$\Ll$}$\;
         $\Ll.\FuncCall{UsedBy}{$\i, \tmin=\i.\tdispatch$}$\;
      }
   }
   \Comment{Dependencies}
   \ForEach{$\var{loc} \in \var{i}.\var{reads}$}{
       $\i.\FuncCall{ConstrainBy}{\loc}$\;
   }
   \Comment{Back-end}
   \ForEach{$\res \in \i.\var{resources}$}{
       $\i.\FuncCall{ConstrainBy}{\res}$\;
       $\res.\FuncCall{UsedBy}{$\i, \tmin=\i.\tdispatch$}$\;
   }
   \Comment{Execution of $\i$}
   $\i.\tstart\assign\i.\tavail$\;
   $\i.\tend \assign \i.\tstart + \i.\var{latency}$\;
   $\i.\tavail \assign \i.\tend$\;
   \ForEach{$\loc \in \i.\var{writes}$}{ 
       \Comment{Assume register renaming works perfectly}
       \If{$\loc \equiv \typehint{Register}$ or $\loc.\tavail < \i.\tend$}{
           $\loc.\FuncCall{SetBy}{\i}$\;
       }
   }
   \Comment{Critical path}
   $\i' \assign \taintqueue.\var{pop}$\;
   \If{$i'\in \dispatch.\taint$}{
     $\taint[\i'.\var{pc}] \assign \taint[\i'.\var{pc}] + 1$\;
   }
   $\i.\taint \assign \i.\taint \cup \{i\}$\;
 }
}
\end{algorithm}

%% file: src/gus-raw-report.tex
\begin{table}
  \resizebox{\columnwidth}{!}{
  \begin{tabular}{lllllllllllll}
      \toprule
                                                     ASM & L2 & p23 &                LAT/PORTS \\
      \midrule
      \cellcolor{orange!25} mov rdx, qword ptr [rsp - 0x10] & 0\% & \cellcolor{orange!25} 10\% &                     2/p23 \\
                       \cellcolor{orange!25} vmovsd xmm0, qword ptr [rdx + rax] & 0\% & \cellcolor{orange!25} 10\% &                     4/p23 \\
          \cellcolor{orange!25} vaddsd xmm0, xmm0, qword ptr [rdx + rax + 8] & 0\% & \cellcolor{orange!25} 10\% & 4/p016 p01 p015 p0156 p23 \\
          \cellcolor{orange!25} vaddsd xmm0, xmm0, qword ptr [rdx + rax + 0x10] & 0\% & \cellcolor{orange!25} 10\% & 4/p016 p01 p015 p0156 p23 \\
                               vmulsd xmm0, xmm0, xmm1 & 0\% &  0\% &     4/p016 p01 p015 p0156 \\
                               \cellcolor{orange!25} mov rdx, qword ptr [rsp - 0x18] & 0\% & \cellcolor{orange!25} 10\% &                     2/p23 \\
                               vmovsd qword ptr [rdx + rax + 8], xmm0 & 1\% & 0\% &                      4/p4 \\
                               \cellcolor{orange!25} mov rdx, qword ptr [rsp - 0x18] & 0\% & \cellcolor{orange!25} 10\% &                     2/p23 \\
                       \cellcolor{orange!25} vmovsd xmm0, qword ptr [rdx + rax - 8] & 0\% & \cellcolor{orange!25} 10\% &                     4/p23 \\
                \cellcolor{orange!25} vaddsd xmm0, xmm0, qword ptr [rdx + rax] & 0\% & \cellcolor{orange!25} 10\% & 4/p016 p01 p015 p0156 p23 \\
                \cellcolor{orange!25} vaddsd xmm0, xmm0, qword ptr [rdx + rax + 8] & 0\% & \cellcolor{orange!25} 10\% & 4/p016 p01 p015 p0156 p23 \\
                               vmulsd xmm0, xmm0, xmm1 & 0\% &   0\% &     4/p016 p01 p015 p0156 \\
                               \cellcolor{orange!25} mov rdx, qword ptr [rsp - 0x10] & 0\% &   \cellcolor{orange!25} 10\% &                     2/p23 \\
                    vmovsd qword ptr [rdx + rax], xmm0 & 0\% &    0\% &                      4/p4 \\
                                        add rax, 0x18 & 0\% &   0\% &                        1/ \\
                                          cmp rax, rcx & 0\% &  0\% &                   1/p0156 \\
      \bottomrule
      \end{tabular}
  % }
  }
  \caption{Fine grain report analysis of resource usage by instruction by \toolsensitivity{} on a \texttt{Jacobi} kernel (Skylake architecture). Only a subset of resources are depicted. The cells colored in orange indicate the bottleneck port and the corresponding instructions.}
  \label{tab:raw-report} 
  \vspace{-2em}
\end{table}

%% file: src/04-evaluation.tex
%
%-------------------------------------------------------------------------------

\section{Experiments}\label{sec:experiments}

We first demonstrate the simulation capabilities of \toolsensitivity{} by comparing it against a cycle-level simulator 
over a set of numerical kernels.

We then validate the portability and accuracy of \toolsensitivity{} against measurements on several microarchitectures.

\label{ssec:framework}
\subsection{Benchmarking framework}

Our benchmarking framework relies on the benchmark suite
PolyBench~\cite{polybench}, composed of 30 numerical computation kernels.
We combine C loop nest optimizers
--~Pluto~\cite{pluto} and PoCC~\cite{pocc}~-- and \texttt{gcc} to
apply several transformations (tiling, loop fusion, unroll and jam, etc.)
to PolyBench. We evaluate \gus on the resulting variants.

The measurements have mostly been performed using the
Performance Application Programming Interface library (PAPI)~\cite{PAPI}.
On the system where the latter is unavailable (the ARM-based one), we used the Linux \texttt{perf} API. In both
cases, we instrumented the benchmark in order to profile a single section
of code (a function call).

To carry out the measurements, we checked that three conditions were met on
the target machines:
\begin{itemize}
\item Simultaneous Multi-Threading must be deactivated, as it tends to
  increase pressure on the front-end, resulting in overestimated hardware
  counters.
\item Processor frequency must be set to the manufacturer's base frequency,
  as frequency variations change memory-boundness.
\item The benchmark should use huge pages to be less sensitive to TLB misses.
\end{itemize}

\label{ssec:gem5}
\subsection{Validation of accuracy and speed against a cycle-level simulator}

\begin{figure}
\begin{subfigure}{\linewidth}
    \centering\small
    \begin{tabular}{|l|l|p{2.4em}|p{2.7em}|p{2.7em}|p{5.9em}|}
        \hline
        \textbf{Simu} & \textbf{\#benchs}& \textbf{MIPS mean} & \textbf{Cycles MAPE} & \textbf{Cycles} \(\tau_{K}\) & \textbf{Retired µops MAPE} \\ \hline
        \gus & 263 & 1.23 & 14.56\% &  0.92 & 2.44\% \\ \hline
        gem5 & 228 & 0.11 & 87.28\% & 0.84 & 52.63\% \\ \hline
        \end{tabular}
    \caption{}
    \label{fig:overall_stats_gus_vs_gem5}
\end{subfigure}
\begin{subfigure}{0.90\linewidth}
    \includegraphics[width=\columnwidth]{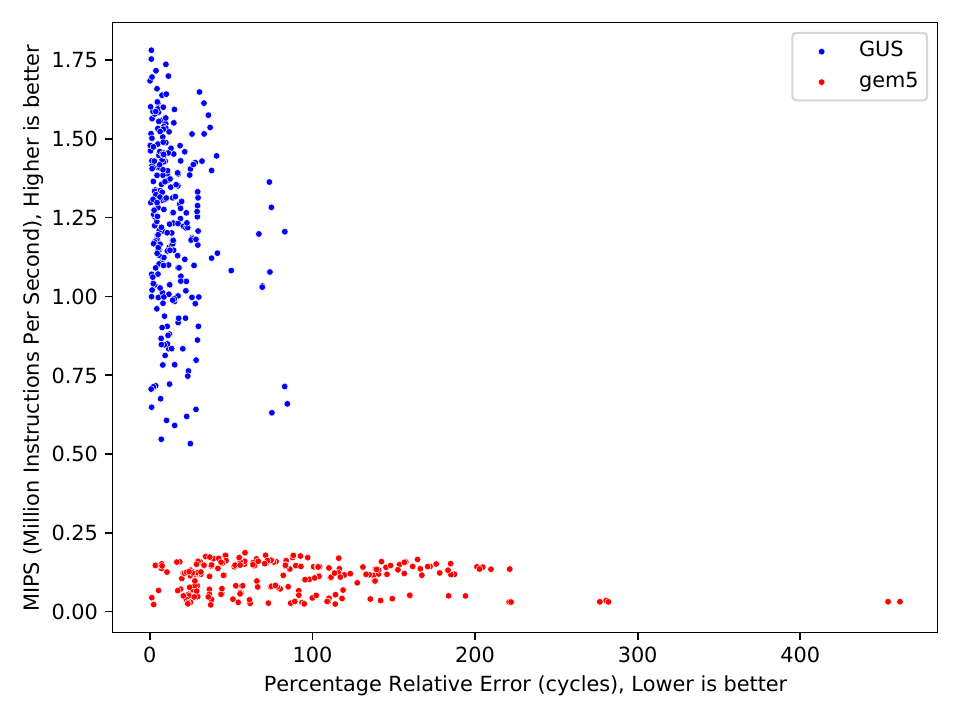}
    \caption{}
    \label{fig:overall_analysis_scatterplot}
\end{subfigure}
	\caption{Statistical comparison (\subref{fig:overall_stats_gus_vs_gem5}) and corresponding error distribution (\subref{fig:overall_analysis_scatterplot}) of accuracy and speed of \gus against the cycle-level simulator gem5.}
\label{fig:overall_analysis_stats_and_scatterplot}
\end{figure}

We present in \autoref{fig:overall_analysis_stats_and_scatterplot} a comparison of the precision and simulation speed of \toolsensitivity{} against a state-of-the-art cycle-level simulator, \texttt{gem5}~\cite{gem5} (v24.0.0.1) 
on a Skylake microarchitecture \footnote{Cascade Lake generation, Intel Xeon Gold 6230R CPU.}.
We use the configuration with the highest level of optimization, \texttt{fast} for \texttt{gem5}, with the \texttt{DerivO3CPU} CPU model and a verbatim description of the processor based on reverse-enginereed characteristics.
Since \texttt{gem5} does not support more recent SIMD extensions than SSE, we limit our \texttt{gcc-13} compiler to generate SSE code.
Over the 410 benchmarks generated by our benchmark framework, 147 had an optimization which did not affect the binary code and were discarded.
An arbitrary timeout of an hour was set for both simulators, that was reached by 35 benchmarks for \texttt{gem5} and none for \gus.

We achieve an order-of-magnitude faster simulation over \texttt{gem5}, with a mean of 1.23 million instructions per second (MIPS) compared to 0.11 MIPS.
We also surpass \texttt{gem5} in accuracy, with a MAPE (Mean absolute percentage error) of 14.56\% compared to 87.28\%, buit also in terms of
Kendall tau~\cite{kendall1938tau} ($\tau_{K} \in [0,1]$, which measures the fraction of
preserved pairwise ordering).

We would like to outline that we focused our implementation efforts for \gus on the relevance of the output and not its speed yet.
In particular, the size of the shadow memory which is the main source of slowdown could be easily reduced.

The low accuracy achieved by \texttt{gem5} can be explained through some of its design characteristics:

\begin{itemize}
    \item Macro-operations are decomposed into micro-operations in \texttt{gem5}. However, the decomposition is imprecise, as it is defined for the whole x86 ISA, and not for the actual microarchitecture. We observe that while \toolsensitivity{} achieves a MAPE of 2.44\% on the number of retired micro-operations, gem5 presents a MAPE of 52.63\%.
    \item No µop-cache is modeled in \texttt{gem5}. This results in a pessimistic front-end model.
\end{itemize}

\subsection{Validation of accuracy against measurements}

We validate the accuracy of \toolsensitivity{} against measurements on several
microarchitectures, as shown in \autoref{tab:validation_microarchitectures}.
This demonstrates its portability and modularity (especially for its performance
model, which is fed in this case by \texttt{uops.info} tables~\cite{uops} for our Intel micro-architectures and by PALMED~\cite{palmed} table for our Cortex-A72). This approach achieves satisfying results on different architectures.

\begin{table}[h]
\centering
\resizebox{\columnwidth}{!}{
\begin{tabular}{|l|l|l|l|}
\hline
\textbf{Microarchitecture} & \textbf{Linux } & \textbf{Release} & \textbf{CPU model} \\ \hline
Maia         & 6.1  & 2015 & ARM Cortex-A72 \\ \hline
Sandy Bridge & 6.11 & 2011 & Intel Core i5-2520m \\ \hline
Skylake X    & 6.11 & 2015 & Intel Core i9-7980XE \\ \hline
Skylake SP   & 5.10 & 2019 & Intel Xeon Gold 6230R \\ \hline
Ice Lake     & 6.11 & 2020 & Intel Core i7-1065G7 \\ \hline
Alder Lake   & 6.11  & 2021 & Intel Core i9-12900k \\ \hline
\end{tabular}
}
\caption{Microarchitectures used for validation of \toolsensitivity{}.}
\label{tab:validation_microarchitectures}
\end{table}

\begin {table}[h]
\centering\small
%\resizebox{\columnwidth}{!}{
\begin{tabular}{|l|l|l|l|}
\hline
\textbf{Microarchitecture} & \textbf{\#benchmarks} & \textbf{MAPE} \\ \hline
Maia         & 289 & 32.13\% \\ \hline
Skylake SP   & 267 & 18.60\%\\ \hline
Skylake X    & 267 & 31.64\% \\ \hline
Ice Lake     & 267 & 24.11\% \\ \hline
Sandy Bridge & 267 & 38.99\% \\ \hline 
Alder Lake   & 267 & 25.01\% \\ \hline
\end{tabular}
%}
\caption{Accuracy of \toolsensitivity{} on different microarchitectures.}
\label{tab:validation_accuracy}
\end{table}

We generate numerous sets of benchmarks for each microarchitecture, with SIMD extensions (AVX2 for x86, NEON for ARM) with a wide range of optimizations as described in \autoref{ssec:framework}.

We would like to stress that our performance model relies solely on automatic resource characterization~\cite{Abel19a,palmed}, which is an orthogonal active research topic.

\subsection{Consistency of sensitivity analysis}

The experimental set-ups where \gus obtains a low MAPE compared to the
measurements indicate that it models well the impact of the program's
instruction mix on the concerned microarchitectures. Thus, we also checked that
the sensitivity gave consistent results. This means that, for a
benchmark B and its optimized version V (\ie, V has a smaller execution time),
then the bottlenecks of B discovered by the sensitivity analysis should appear
equally or less stressed in V than in B. By alleviating the pressure on these
bottlenecks, we have indeed improved the predicted execution time. All the pairs
benchmark/variant of our dataset verify this property.

%% file: src/05-related-works.tex
%-------------------------------------------------------------------------------
\section{Discussion}\label{sec:discussion}
%-------------------------------------------------------------------------------

In this section, we discuss the design choices of \toolsensitivity{} and why 
they are relevant for performance debugging, and how we differentiate from
typical computer architecture simulators. 

\subsection{Trade-offs in the design of \toolsensitivity{}}

Over the years, many microarchitectural simulators~\cite{gem5, marss, sanchez_zsim_2013, sniper2011} have been 
proposed for computer-system architecture research, each with its own set of trade-offs.
Ideally one would like a one-size-fits-all simulator that is both fast and accurate, with a wide applicability to different workloads and fast development time.
However, simulators were designed with different goals in mind (e.g. parallelism, memory hierarchy, etc.), and thus focus on a peculiar dimension of the design space.

Especially, one aspect that is often overlooked with simulators is development cost -- describing a detailed configuration for a microarchitecture down to the µop level can be a daunting task.
In contrast, we propose what we call a \emph{model-based} approach, where we feed a reverse-engineered performance model to a coarse-grained core model to find performance bottlenecks. 
This approach covers less of the design space than a full-fledged simulator, as it sacrifices coverage and some accuracy, but enables finding performance bottlenecks in a reasonable amount of time and effort.

We summarize briefly the major trade-offs of simulation approaches in Table~\ref{tab:simulators} to position \toolsensitivity{} among them. We differentiate between cycle-level simulators (e.g. gem5~\cite{gem5}), cycle-approximate simulators (e.g. Sniper~\cite{sniper2011}, ZSim~\cite{sanchez_zsim_2013}), and model-based simulators (e.g. \toolsensitivity{}).

\begin{table}[h]
\centering
\resizebox{\columnwidth}{!}{%
\begin{tabular}{|l|l|l|l|l|l|}
\hline
\textbf{Type} & \textbf{Dev} & \textbf{Accuracy} & \textbf{Speed} & \textbf{Coverage} \\ \hline
Cycle-level   & Slow  & High          & Low        & High             \\ \hline
Cycle-approximate & Slow   & Medium   & High        & High             \\ \hline
% insert separator
\hline
\textbf{Model-based}  & \textbf{Fast}   & \textbf{Medium}           & \textbf{High}       & \textbf{Low}              \\ \hline
\end{tabular}
}
\caption{Comparison of different approaches to microarchitectural simulation.}
\label{tab:simulators}
\end{table}

% Overall, keeping in mind performance debugging as the main goal, we found several limitations of gem5~\cite{gem5} that hinder its applicability to performance debugging tasks in HPC applications:

% \begin{itemize}
%     \item Inaccurate decomposition to µops~\cite{harmful}, as shown in \alban{UPDATE THIS} due to a fixed-decomposition per ISA created mostly for functional correctness.
%     \item Outdated support of SIMD instructions, which are crucial for performance debugging of HPC applications (e.g. AVX2).
% \end{itemize}

%% file: src/06-conclusion.tex
%-------------------------------------------------------------------------------
\section{Conclusion}\label{sec:conclusion}
%-------------------------------------------------------------------------------

Identifying performance bottlenecks in programs that need to make the most of the architecture is an increasingly critical task. 
We present micro-architectural mechanisms on which program performance depends, as well as the metrics and analysis methods that have been designed to guide optimization. 
We propose to generalize bottleneck analysis by means of microarchitectural sensitivity and constraint causality. 
Sensitivity analysis aims to automatically discover which resources influence a program's overall performance by successively accelerating each of them and observing the overall speedup generated by this variation.
Causality analysis aims to propagate timing constraints through the simulation process to pinpoint how each instruction contribute to the overall execution time.
We present \gus, a dynamic code analyzer that implements sensitivity and causality analysis. 
We evaluate both its performance model and bottleneck analysis algorithm on a set of benchmarks generated from the PolyBench benchmark suite, and highlight its strengths and limitations compared with existing methods based on hardware counters or performance models. 
\gus's performance model achieves state-of-the-art accuracy in throughput estimation over our experimental harness. 
As for our bottleneck analysis, it allows us to enhance the results of existing approaches, especially where saturation is not sufficient to identify the one that limits parallelism.